\input{aipcheck}

\documentclass[
    ,final            % use final for the camera ready runs
  ]
  {aipproc}

\layoutstyle{6x9}

\usepackage{graphicx}
\newcommand{\be}{\begin{eqnarray}}
\newcommand{\ee}{\end{eqnarray}}
\newcommand{\ket}[1]{\vert\,{#1}\rangle}
\newcommand{\eps}{\epsilon}

\begin{document}

\title{Diffraction Patterns in Deeply Virtual Compton Scattering}


\author{Asmita Mukherjee}{
  address={Physics Department, Indian Institute of Technology Bombay,\\ 
Powai, Mumbai 400076, India}
}

\begin{abstract}
We report on a calculation to show that the Fourier transform of 
the Deeply Virtual Compton Scattering (DVCS) amplitude with respect to the 
skewness variable $\zeta$ at fixed invariant momentum transfer squared $t$
gives results that are
analogous to the diffractive scattering of a wave in optics. 
As a specific example, we utilize the quantum fluctuations of a fermion 
state at one loop in QED to obtain the behavior of the DVCS amplitude for 
electron-photon scattering. We then simulate the wavefunctions for a hadron 
by differentiating the above LFWFs with respect to $M^2$ and study the 
corresponding DVCS amplitudes in light-front longitudinal space. 
\end{abstract}

\maketitle
%%%%%%%%%%%%%%%%%%%%%%%%%%%%%%%%%%%%%%%%%%%%%%%%%%%%%%%%%%%%%%%%%%%
\section{Introduction}
Measurements of Deeply Virtual Compton Scattering (DVCS) cross sections 
with specific proton
and photon polarizations can provide comprehensive probes of the spin as
well as spatial structure of the proton at the most fundamental level of QCD.
The DVCS process involves off-forward hadronic matrix elements of light-front 
bilocal currents. The DVCS quark
matrix elements can  be computed  from the off-diagonal overlap of the 
boost-invariant light-front Fock state wavefunctions (LFWFs) of the target
hadron~\cite{overlap2,overlap1}.  The longitudinal momentum transfer to the
target hadron is given by the `skewness' variable $\zeta={Q^2\over 2 p\cdot q}.$

Fourier transform (FT) of the off-forward or generalized 
parton distributions (GPDs) with respect to the transverse momentum 
transfer $\Delta^\perp$ at zero skewness and fixed
longitudinal momentum fraction $x$ gives the parton distributions in the 
impact parameter ($b_\perp$) space \cite{bur1,soper}. Also, 
a 3D picture of the proton has been proposed in \cite{wigner} in terms of 
a Wigner  distribution for the relativistic  quarks and  gluons inside the 
proton. 

We introduce a  coordinate $b$ conjugate to the
momentum transfer $\Delta$ such that $ b \cdot \Delta = \frac{1}{2} b^+
\Delta^- + \frac{1}{2} b^- \Delta^+ - b_\perp \cdot \Delta_\perp$. Note that
$ \frac{1}{2} b^- \Delta^+ = \frac{1}{2} b^- P^+ \zeta = \sigma \zeta $
where we have defined the boost invariant variable $\sigma$
which is an $` $impact parameter' in the longitudinal coordinate space.
The Fourier transform
of  the  DVCS amplitude with respect to $\zeta$ allows one to determine the
longitudinal structure of the target hadron in terms of the variable
$\sigma$.

%%%%%%%%%%%%%%%%%%%%%%%%%%%%%%%%%%%%%%%%%%%%%%%%%%%%%%%%%%%%%%%%%%%%%%%%%%%% 
\section{DVCS amplitude in $\sigma$ space}

In order to illustrate  our general framework, we will present an
explicit calculation of the  $\sigma$ transform of  virtual Compton scattering
one-loop order \cite{drell}. The GPDs in 
impact parameter space have been calculated in this model \cite{dip}.
One can generalize this analysis
by assigning a mass $M$ to the external electrons and a different
mass $m$ to the internal electron lines and a mass $\lambda$ to the
internal photon lines with $M < m + \lambda$ for stability. 

The light-front Fock state wavefunctions corresponding to the
quantum fluctuations of a physical electron can be systematically
evaluated in QED perturbation theory. The state is expanded in
Fock space and there
are contributions from $\ket{e^- \gamma}$ and $\ket{e^- e^- e^+}$,
in addition to renormalizing the one-electron state at one loop 
level \cite{overlap2}. We use the handbag approximation. In the domain 
$\zeta <x <1$, there are 
diagonal $2 \to 2$ overlap contributions, \cite{overlap2}.
The GPDs $H_{(2\to 2)}(x,\zeta,t)$ and $E_{(2\to
2)}(x,\zeta,t)$ are zero in the domain $\zeta-1 < x < 0$, which
corresponds to emission and reabsorption of an $e^+$ from a physical
electron. The contributions in the domain, $0 < x < \zeta$ come
from  overlaps of three-particle and one-particle LFWFs \cite{overlap2}.
A summary of our main results is given in \cite{letter} and details
of the calculation are given in \cite{article}.

\begin{figure}[t]
\includegraphics[width=6.2cm,height=6.2cm,clip]{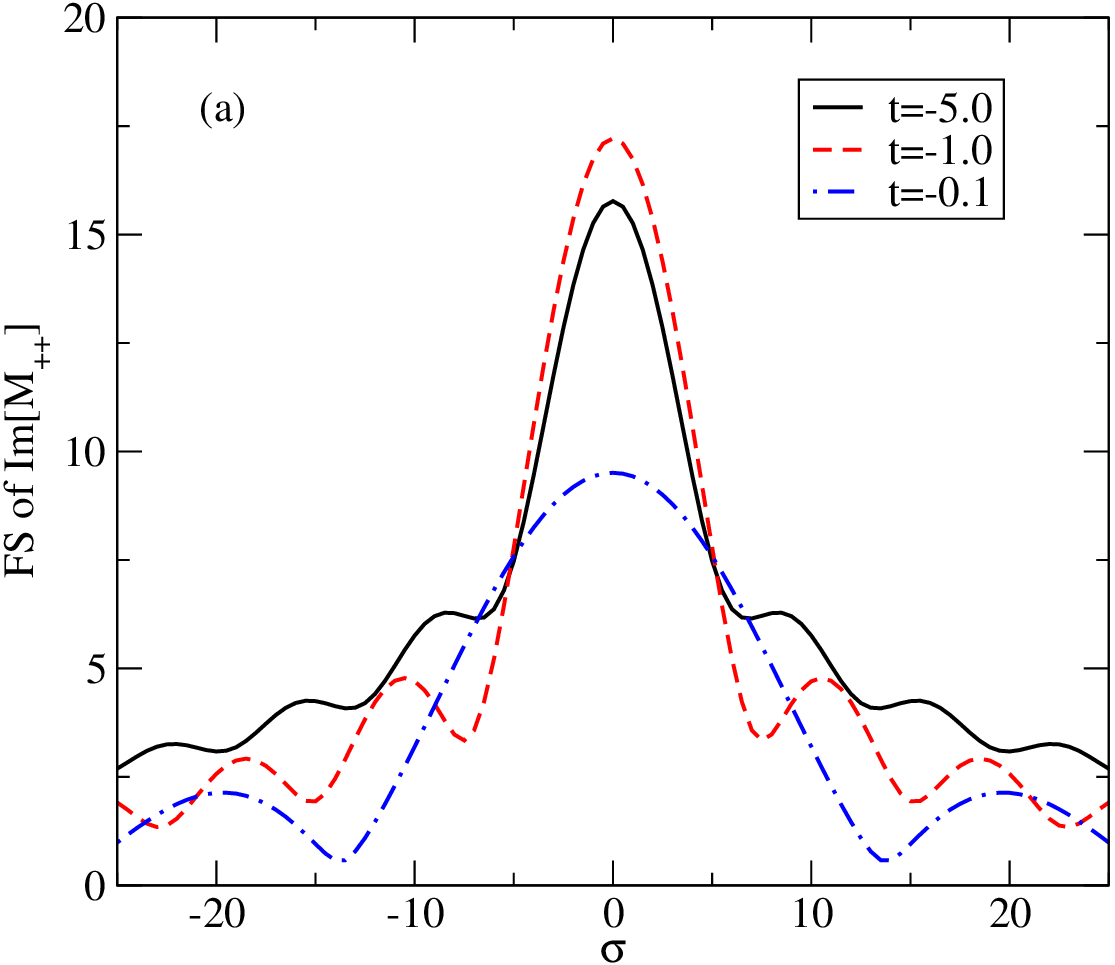}%
\hspace{0.2cm}%
\includegraphics[width=6.2cm,height=6.2cm,clip]{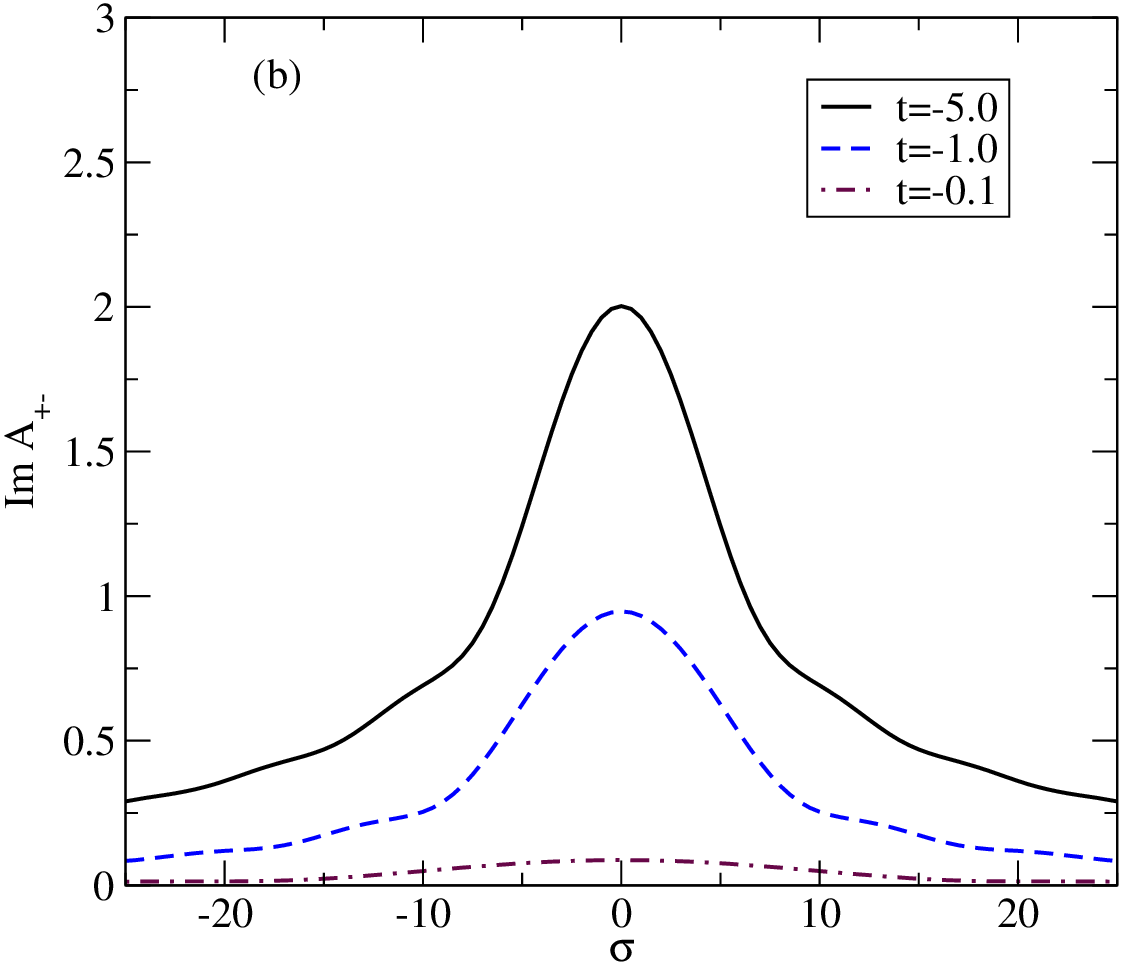}
\caption{\label{fig1} Fourier spectrum of the imaginary part of
the DVCS amplitude of an electron vs. $\sigma$
for $M=0.51$ MeV, $m=0.5$ MeV,
$\lambda=0.02$  MeV, (a) when the electron helicity is not flipped;
(b) when the helicity is flipped.
The parameter $t$ is in ${\mathrm{MeV}^2}$.}
\end{figure}

The off-forward matrix elements for both $2 \to 2$ and $3 \to 1$ overlaps 
are calculated using the analytic form of the LFWFs. There can be two types 
of contributions, one in which the helicity of the target electron is not 
flipped, in this case both the GPDs $H$ and $E$ will contribute; in the 
other contribution the helicity is flipped and only $E$ contributes. 
There are real and imaginary parts of the DVCS amplitude, both of which 
can be accessed in the experiment. The imaginary part receives contribution 
only when $x=\zeta$, because of the delta function coming from the propagator 
in the hard part of the amplitude. The other values of $x$ contribute to the 
real part.  The GPDs are continuous at $x=\zeta$, and a principal value 
prescription is used for the propagator in the real  part. However, 
there are potential divergences coming from $x \to 1$ region 
as well as small $x$, $\zeta$ region when we take a Fourier transform (FT) 
in $\zeta$. We use a cutoff scheme discussed in \cite{article}. 
The FT of the DVCS amplitude are obtained as :
\be
A_{\lambda, \lambda'} (\sigma, t) =
{1\over 2 \pi } \int_{\eps_2}^{1-\eps_2} d\zeta ~ e^{{i}
\sigma \zeta }~ M_{\lambda,\lambda'}
(\zeta, \Delta_\perp);
\ee
where $M_{\lambda,\lambda'}$ is the DVCS amplitude, $\lambda$ ($\lambda'$) 
are the  helicities of the initial (final) electron. 
$\sigma = {1\over 2} P^+ b^-$ is the (boost invariant) longitudinal 
distance on  the light-cone and the FTs are performed at a fixed invariant 
momentum transfer squared $-t$.
We have imposed cutoffs $\eps_2=0.002$ for the numerical calculation.
Fourier transforms have been performed by numerically calculating the
Fourier sine and cosine transforms and then calculating the resultant by
squaring them, adding and taking the square root, thereby yielding the
Fourier Spectrum (FS).

\begin{figure}[t]
\centering
\includegraphics[width=6.2cm,height=6.2cm,clip]{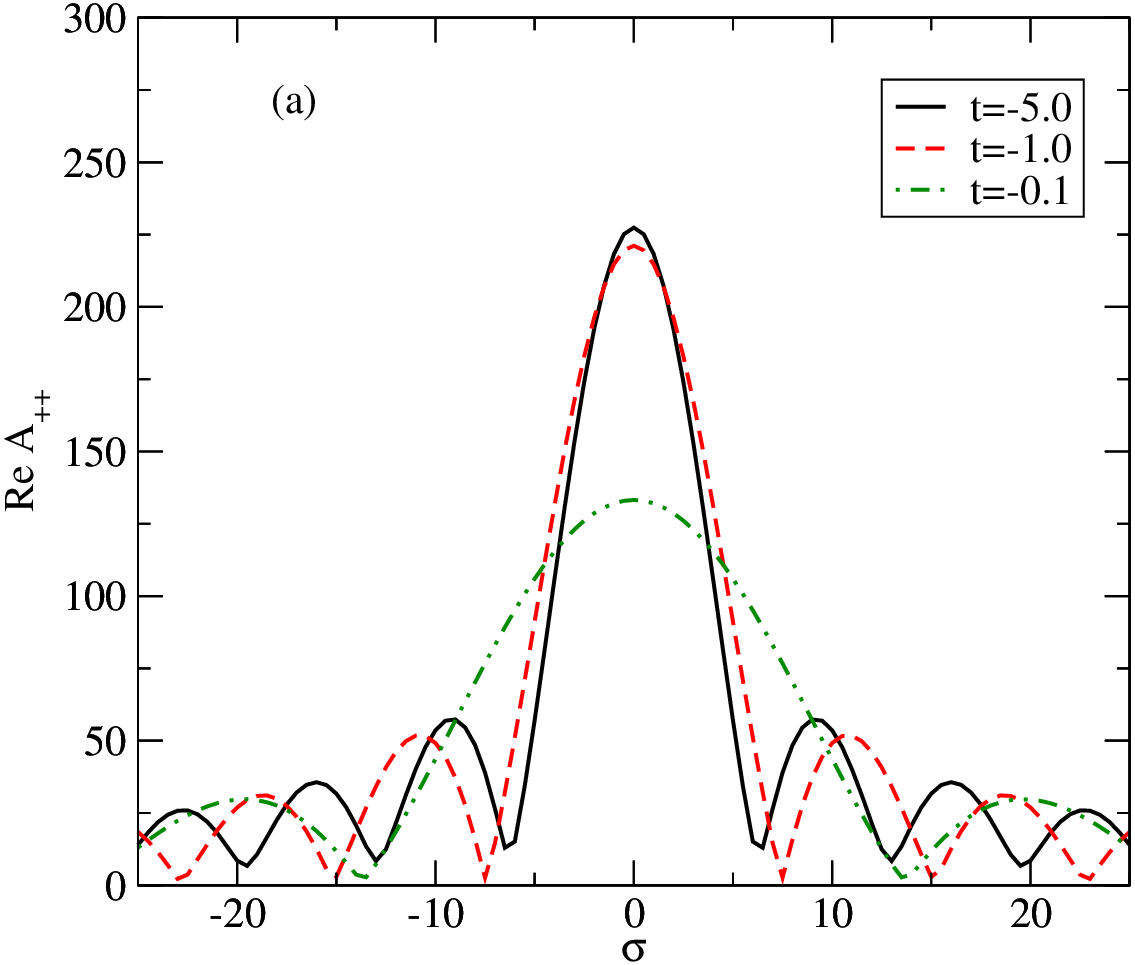}%
\hspace{0.2cm}%
\includegraphics[width=6.2cm,height=6.2cm,clip]{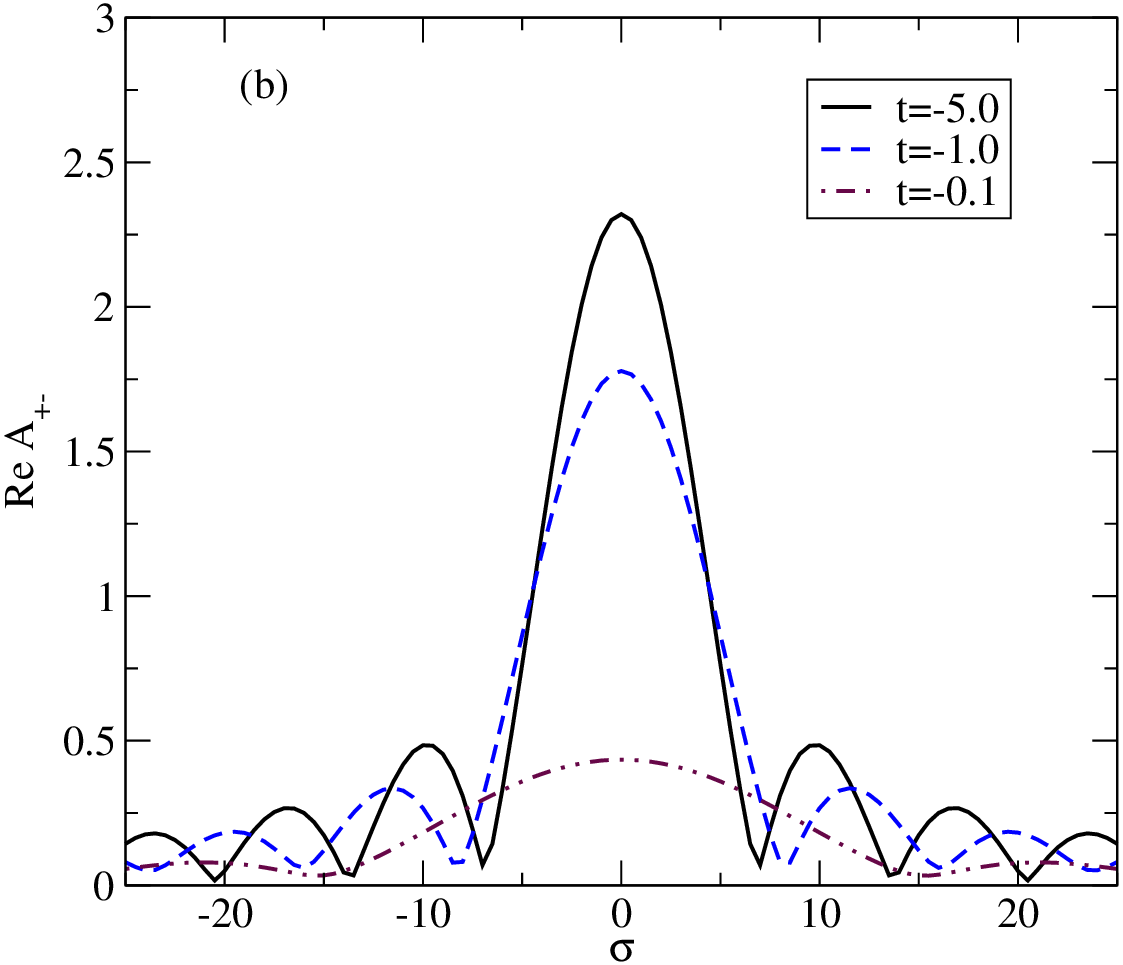}
%\end{minipage}%
\caption{\label{fig2} Fourier spectrum of the real part of
the DVCS amplitude of an electron vs. $\sigma$
for $M=0.51$ MeV, $m=0.5$ MeV,
$\lambda=0.02$  MeV, (a) when the electron helicity is  not flipped;
(b) when the helicity is flipped.
The parameter $t$ is in ${\mathrm{MeV}^2}$.}
\end{figure}
   
In Fig. 1 and 2 respectively, we have shown the FS of the imaginary 
and real parts of the DVCS  amplitude. Apart from the imaginary part 
of the helicity-flip amplitude, they show a diffraction pattern in 
$ \sigma $. The helicity non-flip part depends on the scale $Q$, which 
we took to be $10$ MeV.  

In the $2-$ and $3$-body LFWFs, the bound-state mass squared $M^2$ appears
in the denominator.  Differentiation of the LFWFs with respect to $M^2$
increases the fall-off of the wavefunctions near the end points $x=0,1$
and mimics the hadronic wavefunctions. 
One has to note that differentiation of the single particle wave function
yields zero and thus there is no $3-1$ overlap contribution to the DVCS
amplitude in this hadron model. Also, the imaginary part of the amplitude 
vanishes in this model. We show the amplitudes in Fig. 3.
We have also computed the DVCS amplitude and the
corresponding diffraction pattern in
$\sigma$ using the AdS/QCD framework \cite{article}. This model provides a 
useful  first approximation to hadron wavefunctions which has confinement 
at large distances and conformal behavior at short distances \cite{tera}.

\begin{figure}
\centering
\includegraphics[width=6.2cm,height=6.2cm,clip]{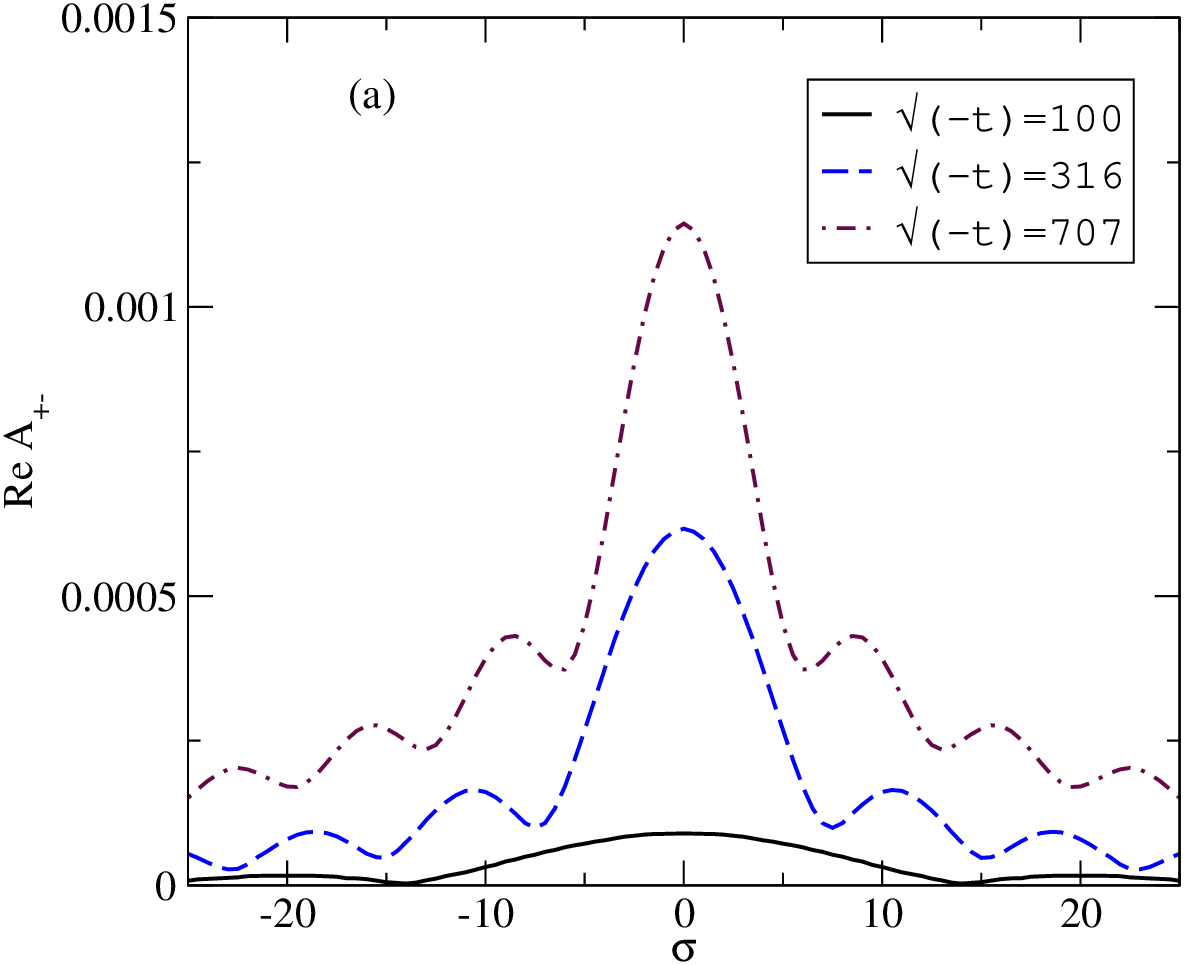}%
\hspace{0.2cm}%
\includegraphics[width=6.2cm,height=6.2cm,clip]{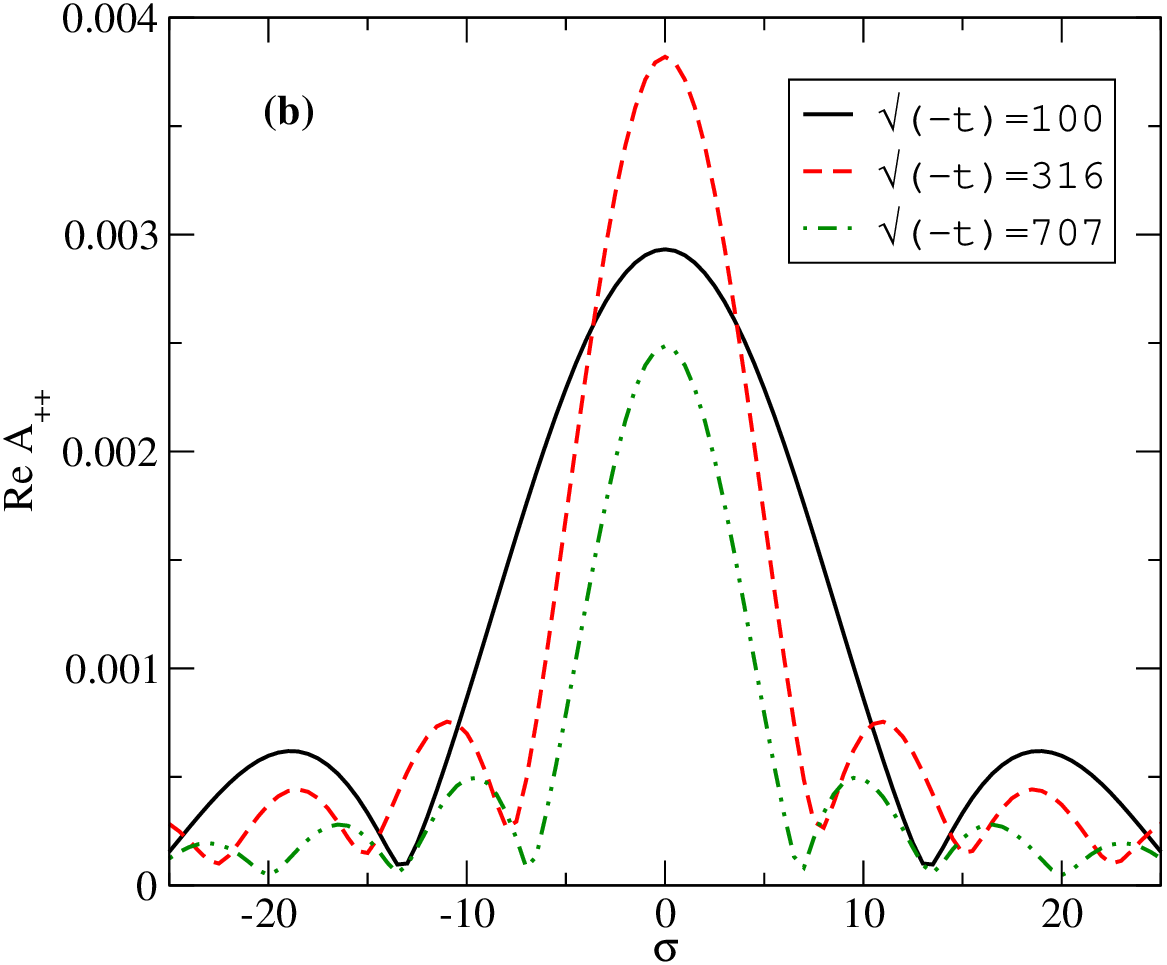}
%\end{minipage}%
\caption{\label{fig3} Fourier spectrum of the real part of the 
DVCS amplitude for the simulated meson-like bound state.
The parameters are $M=150, m=\lambda=300$ MeV.
(a) Helicity flip amplitude (b) helicity non-flip amplitude.
The parameter $t$ is in ${\mathrm{MeV}^2}$.}
\end{figure}

From the plots, we propose an optics analog of the behavior of the
Fourier Spectrum of the DVCS amplitude. The diffractive
patterns in $\sigma$ sharpen and the positions of the first minima
typically move in with increasing momentum transfer. For fixed $-t$, 
higher minima appear at
positions which are integral multiples of the lowest minimum, similar 
to the diffraction pattern in optics. Thus the
invariant longitudinal size of the parton distribution becomes
longer and the shape of the conjugate light-cone momentum
distribution becomes narrower with increasing $\mid t \mid$.
If one Fourier transforms in  $\zeta$ at fixed $\Delta_\perp$ and 
then Fourier transforms the change in transverse momentum $\Delta_\perp$
to impact space $b_\perp$ \cite{bur1}, then one would have the
analog of a three-dimensional scattering center. In this sense,
scattering photons in DVCS provides the complete Lorentz-invariant
light front coordinate space structure  of a hadron.

\begin{theacknowledgments}
This work has been done in collaboration with S. J. Brodsky, D. Chakrabarti, 
A. Harindranath and J. P. Vary. I  thank  the organizers of Spin 2006, Kyoto
for the invitation and support and IIT Bombay for support.  
\end{theacknowledgments}

%%%%%%%%%%%%%%%%%%%%%%%%%%%%%%%%%%%%%%%%%%%%%%%%
%% The bibliography can be prepared using the BibTeX program or
%% manually.
%%
%% The code below assumes that BibTeX is used.  If the bibliography is
%% produced without BibTeX comment out the following lines and see the
%% aipguide.pdf for further information.
%%
%% For your convenience a manually coded example is appended
%% after the \end{document}
%%%%%%%%%%%%%%%%%%%%%%%%%%%%%%%%%%%%%%%%%%%%%%%%

%%%%%%%%%%%%%%%%%%%%%%%%%%%%%%%%%%%%%%%%%%%%%%%%
%% You may have to change the BibTeX style below, depending on your
%% setup or preferences.
%%
%%
%% For The AIP proceedings layouts use either
%%%%%%%%%%%%%%%%%%%%%%%%%%%%%%%%%%%%%%%%%%%%

\bibliographystyle{aipproc}   % if natbib is available
%\bibliographystyle{aipprocl} % if natbib is missing

%%%%%%%%%%%%%%%%%%%%%%%%%%%%%%%%%%%%%%%%%%%
%% You probably want to use your own bibtex database here
%%%%%%%%%%%%%%%%%%%%%%%%%%%%%%%%%%%%%%%%%%%
%\bibliography{sample}

%%%%%%%%%%%%%%%%%%%%%%%%%%%%%%%%%%%%%%%%%%%
%% Just a reminder that you may have to run bibtex
%% All of it up to \end{document} can be removed
%% if you don't like the warning.
%%%%%%%%%%%%%%%%%%%%%%%%%%%%%%%%%%%%%%%%%%%
%\IfFileExists{\jobname.bbl}{}
% {\typeout{}
%  \typeout{******************************************}
%  \typeout{** Please run "bibtex \jobname" to optain}
%  \typeout{** the bibliography and then re-run LaTeX}
%  \typeout{** twice to fix the references!}
%  \typeout{******************************************}
%  \typeout{}
% }

%\end{document}

%%%%%%%%%%%%%%%%%%%%%%%%%%%%%%%%%%%%%%%%%%%
%% The following lines show an example how to produce a bibliography
%% without the help of the BibTeX program. This could be used instead
%% of the above.
%%%%%%%%%%%%%%%%%%%%%%%%%%%%%%%%%%%%%%%%%%%

%\endinput
\end{document}